\documentstyle[PASJadd,epsf]{PASJ95}
\draft

\begin{document}

\title{Gas, Iron and Gravitational Mass in Galaxy Clusters:\\
The General Lack of Cluster Evolution at $z < 1.0$}

\author{Hironori {\sc Matsumoto},$^1$ Takeshi Go {\sc Tsuru},$^2$\\
Yasushi {\sc Fukazawa},$^3$ Makoto {\sc Hattori},$^4$ and 
David S. {\sc Davis}$^1$
\\[12pt]
$^1$ {\it Center for Space Research, Massachusetts Institute of Technology,}\\
{\it 77 Massachusetts Avenue, Cambridge, MA02139-4307, USA}\\
{\it E-mail(HM): matumoto@space.mit.edu}\\
$^2$ {\it Department of Physics, Faculty of Science, Kyoto University,}\\
{\it Sakyo-ku, Kyoto 606-8502}\\
$^3$ {\it Department of Physics, Graduate School of Science, The University of Tokyo,}\\
{\it 7-3-1 Hongou, Bunkyou-ku, Tokyo 113-0033}\\
$^4$ {\it Astronomical Institute, Tohoku  University,
Aoba Aramaki, Sendai 980-8578}
}

\abst{ 

We have analyzed the ASCA data of 29 nearby clusters of
galaxies systematically, and obtained temperatures, iron
abundances, and X-ray luminosities of their intracluster
medium (ICM). We also estimate ICM mass using the $\beta$
model, and then evaluate iron mass contained in the ICM and
derive the total gravitating mass. This gives the largest
and most homogeneous information about the ICM derived only
by the ASCA data.  We compare these values with those of
distant clusters whose temperatures, abundances, and
luminosities were also measured with ASCA, and find no clear
evidence of evolution for the clusters at $z<1.0$. Only the
most distant cluster at $z=1.0$, AXJ2019.3+1127, has
anomalously high iron abundance, but its iron mass in the
ICM may be among normal values for the other clusters,
because the ICM mass may be smaller than the other
clusters. This may suggest a hint of evolution of clusters
at $z \sim 1.0$.}

\kword{Galaxies: clusters of --- Cosmology --- X-rays: spectra}

\maketitle
\thispagestyle{headings}

\section{Introduction}

Clusters of galaxies are the largest bound systems in the
universe and a major fraction of their visible mass is X-ray
emitting hot gas which is known as the intracluster medium
(ICM).  The physical conditions of the ICM are largely
determined by the nature of the dynamical and chemical
evolution and the dark matter distribution of
clusters. Accordingly, X-ray studies for structures of
clusters and their evolution provide key information for
cosmology.

The main purpose of this paper is to present a large and
homogeneous dataset on the temperature, metallicity,
luminosity, and density distribution of the ICM in the
nearby clusters derived only from the ASCA data (Tanaka et
al.\ 1994).  We used the $\beta$ model to derive the ICM
distribution. We then estimated the ICM mass ($M_{\rm
gas}$), the iron mass ($M_{\rm Fe}$) in the ICM, and the
total gravitational mass ($M_{\rm tot}$) from our results.

Tsuru et al.\ (1996) and Mushotzky and Scharf (1997)
compiled a catalogue using the ASCA data of distant clusters
(mostly $0.1<z<0.6$), and compared it with the data from
nearby clusters obtained with the Einstein and ROSAT
observatories (e.g. David et al.\ 1993).  They suggested
that no systematic differences exist in the X-ray luminosity
($L_{\rm X}$) -- temperature ($kT$) relation between distant
clusters and nearby clusters. Mushotzky and Loewenstein
(1997) compared iron abundances ($A_{\rm Fe}$) in the ICM of
nearby clusters with those of distant clusters, and found no
evidence for evolution of $A_{\rm Fe}$ at $z \ltsim
0.3$. Tsuru et al.\ (1996) along with Mushotzky and
Loewenstein (1997) studied the $kT$ -- $A_{\rm Fe}$
relation, and concluded that there are no differences
between nearby and distant clusters.  However, these studies
are based on comparison of the results from different
instruments (ASCA, ROSAT, Ginga and Einstein), which should
be treated carefully unless cross-calibration between these
instruments are fully performed.  Thus, the other purpose of
this paper is to compare the temperatures, metal abundances,
and luminosities of nearby clusters with those of distant
clusters using only the ASCA data.  We also study the
evolution of $M_{\rm gas}$, $M_{\rm Fe}$, and $M_{\rm tot}$
by comparing these values for nearby clusters with those of
distant clusters in the literature.

Throughout this paper, we assume $H_0$ = 50 km/s/Mpc and
$q_0$ = 0.5. All errors used in this paper are at 90 \%
confident level for one interesting parameter except for
AXJ2019. The errors for AXJ2019 are at 1 $\sigma$ level.

\section{Data Analysis and Results of Nearby Clusters}

We selected the ASCA sample of clusters with redshifts less
than 0.1 using the following requirement: the cluster must
be reasonably extended so that it is spatially resolved with
ASCA, the X-ray flux is high so that we can constrain the
iron abundance, and its morphology is nearly symmetric to
exclude any merger or dynamic effects. We made an effort to
include clusters with a wide range of gas
temperatures. Finally, there are 29 clusters in our sample
which are listed in table 1.  The average redshift is 0.032.
The ASCA data were screened with the standard selection
criteria to exclude such data as affected by the South
Atlantic Anomaly, Earth occultation, and regions of low
geomagnetic rigidity (Fukazawa 1997).

The GIS spectra for each of the nearby clusters was
accumulated from a circular region with a radius of 15
arcmin from the cluster center.  We did not use the SIS
data, because we did not use them for the imaging analysis
as discussed below. We believe that the GIS data alone are
sufficient to constrain the temperature and iron abundance
of the clusters, because the temperatures of our sample
clusters are higher than 2 keV and the GIS has higher
sensitivity than the SIS in the energy band above 2
keV. Some of the nearby clusters are known to exhibit
two-temperature thermal spectra at their central regions
(Fukazawa 1997).  However, it is rather difficult to
distinguish between a single-temperature spectrum and a
multi-temperature spectrum if the total count of the
spectrum is low. Since the fluxes of distant clusters are
inevitably low, it is difficult to find the
multi-temperature structure in the X-ray spectrum of the
distant clusters.  It was confirmed that the X-ray spectra
of distant clusters at redshifts higher than 0.1 were well
described with the single-temperature thin thermal plasma
model (Mushotzky, Loewenstein 1997; Mushotzky, Scharf 1997).
Since one of the main purposes of this paper is to
systematically compare the ASCA results of the nearby
clusters with those of the distant clusters, we fitted all
the GIS spectra of the nearby clusters with the
single-temperature model (Masai 1984) with a free absorption
($N_{\rm H}$). In this fitting, we fixed abundance ratios
between various elements to the solar ratio ((Fe/H)$_\solar$
= $4.68\times10^{-5}$; Anders, Grevesse 1989). Since our
sample of nearby clusters have temperatures higher than 2
keV and we used only the GIS data, the metal abundances were
mainly determined by the iron K line.  Therefore, we regard
the determined abundances as those of iron.  We show the
results in table 1.  As for the X-ray luminosity, we limited
the energy range to be 2 -- 10 keV in order to exclude
possible contamination from the cool component found in some
clusters.

We then analyzed the X-ray surface brightness of the nearby
clusters using the GIS data. We did not use the SIS data
because the clusters extend beyond its field of view. We
made azimuthally averaged radial profiles of X-ray counts
typically in the 1.8 -- 7.1 keV band to minimize
contribution from the central cool component. We fitted the
profile with the $\beta$ model taking account of the complex
PSF effects of ASCA (e.g. Takahashi et al.\ 1995). The
$\beta$ model, $S(r)$, is given by
\begin{equation}
S(r)=S(0)\left[1+\left(\frac{r}{r_{\rm core}}\right)^2\right]^{-3\beta+\frac{1}{2}},
\end{equation}
where $r$ is the radius and $r_{\rm core}$ is the core
radius. The best-fit $\beta$ profile can be converted to the
proton density profile of the ICM, $n(r)$, which is
expressed as
\begin{equation}
n(r)=n(0)\left[1+\left(\frac{r}{r_{\rm core}}\right)^2\right]^{-\frac{3}{2}\beta}
\end{equation}
(e.g. Sarazin 1986). We then integrated the density profile
to estimate the ICM mass within 1 Mpc from the center. Our
results are consistent with those derived by the Einstein or
ROSAT data (e.g. White et al.\ 1997; Mohr et al.\ 1999).
The iron mass included in the ICM within 1 Mpc from the
center was estimated by using the best-fit iron abundances
in table 1. If we assume hydrostatic equilibrium for the ICM
and the isothermal ICM distribution of the $\beta$ model,
the total gravitational mass within a radius $R$, $M(<R)$,
is expressed as
\begin{equation}
M(<R) = 3\beta\frac{kTR}{{\mu}{m_{\rm p}}G}\frac{(R/r_{\rm core})^2}{1+(R/r_{\rm core})^2},
\end{equation}
where $G$ is the gravitational constant, $\mu$ is the mean
molecular weight (we assume $\mu$ = 0.6), and $m_{\rm p}$ is
the mass of a proton. We estimate the total mass within the
1 Mpc radius using the equation (3), and the results are
shown in table 2. Further details of the observations and
analysis of the ASCA data of the nearby clusters are found
in Fukazawa (1997).

\section{Discussion}

In this section, we compare our results with other distant
clusters in the literature. Most of our sample of distant
clusters are from Mushotzky, Scharf (1997) and Mushotzky,
Loewenstein (1997). Therefore, we should note that the sample
has a strong bias against clusters of $L_X<10^{45}$ erg/s,
as noted by Mushotzky, Scharf (1997). The temperatures, iron
abundances, and X-ray luminosities of the distant clusters
were determined only by the ASCA data.  Unfortunately, the
ASCA imaging quality is insufficient to extract the
morphological parameters for most of the distant clusters;
hence, we use the ICM and total masses derived with the
Einstein or ROSAT data from the literature. Otherwise, we
used $\beta$ model parameters derived with the Einstein or
ROSAT data in the literature, and we calculated masses by
using them. The iron mass in the ICM was calculated by
combining the ICM mass derived as mentioned above and the
iron abundance determined with ASCA. 

Since the ASCA imaging quality is worse than Einstein and
ROSAT, the parameters listed in table 2 have generally
larger errors than those determined by the Einstein and
ROSAT data in other work (White et al.\ 1997; Mohr et al.\
1999). However, they are consistent with each
other. Therefore, systematic calibration errors between
different instruments for the morphological parameters are
less serious than those of spectroscopic parameters, in
particular, temperatures and abundances. Our sample for the
distant clusters are listed in table 3 with all relevant
parameters. We should note that the errors for AXJ2019 is at
1 $\sigma$ level while the others are at 90 \% level,
because the original paper (Hattori et al.\ 1997) shows only
the 1 $\sigma$ errors.

We show the $kT$ -- $L_{\rm X}$ relation in figure 1. There
is no significant difference between $z<0.1$ and
$0.1<z<1.0$.  Although the clusters at $0.1<z<1.0$ show a
somewhat flatter slope than the nearby clusters, this is
probably due to the selection bias for the distant clusters
as already noted by Mushotzky, Scharf (1997). The most
distant cluster, AXJ2019 denoted by a star in figure 1, is
also consistent with the other clusters.

Figure 2 shows the $kT$ -- $M_{\rm gas}$ relation. No X-ray
emission from the ICM in AXJ2019 was detected beyond 0.5 Mpc
from the center (Hattori et al.\ 1997). Therefore, we used
the best-fit parameters of the $\beta$ model and
extrapolated to 1 Mpc from the center, and both of these are
plotted in figure 2. We found no difference between the
clusters at $z<0.1$ and $0.1<z<1.0$. The most distant
cluster, AXJ2019, is marginally consistent with the other
clusters taking its large errors into account.  However, the
best-fit values may suggest that the cluster has less gas
mass than the other clusters of similar temperatures.  The
reason why AXJ2019 may have such a low gas mass in spite of
its normal luminosity is that the $\beta$ of AXJ2019 ($\beta
\sim 0.9$) is rather larger than the other clusters ($\beta
\sim 0.6$).

Figure 3 shows the $kT$ -- $A_{\rm Fe}$ relation.  We see no
clear differences between the clusters at $z<0.1$ and
$0.1<z<1.0$. However, the most distant cluster has an
extremely large best-fit abundance. Most of the clusters
having dominant galaxies at their centers show cool
components in their X-ray spectra from the central regions.
The iron abundance of the cool component is often higher
than the surrounding ICM (e.g. Fukazawa 1997). Some clusters
in our nearby sample have dominant galaxies, and then they
tend to show low-temperatures and high-metallicities in our
analysis as well.  We believe this can explain the tendency
that the cool clusters in figure 3 have larger abundances
than the hot clusters. Also, Fukazawa et al.\ (1998)
analyzed the X-ray spectra of our sample, excluding the
central regions, and found no evidence for the temperature
dependence of the iron abundance, which is consistent with
our results.

We show the $kT$ -- $M_{\rm Fe}$ relation in figure 4. We
also plotted two points for AXJ2019 as described above.
There is no clear difference between the clusters at
$z<0.1$ and $0.1<z<1.0$. Furthermore AXJ2019 is also
consistent with the other clusters, although its ICM mass may be
extremely low. This is because the extremely high abundance
compensates for the low gas mass.

Figure 5 shows the $kT$ -- $M_{\rm tot}$ relation.  There is
no clear difference between the clusters at $z<0.1$ and
$0.1<z<1.0$. The most distant cluster AXJ2019 is consistent
with the other clusters, while the ICM may be less massive
than the other clusters.  This may indicate that the
formation of the gas halo and the dark matter halo in
cluster is not a simultaneous process, and the gas
accumulation process continues after the dark matter halo
formation is completed.

These data show that there are no differences between the
clusters at $z<0.1$ and $0.1<z<1.0$. The most distant
cluster at $z=1.0$, AXJ2019, may be different from the other
clusters at $z<1.0$ in terms of the $kT$ -- $M_{\rm gas}$
and $kT$ -- $A_{\rm Fe}$ relations, although we should note
that there is still room to allow AXJ2019 to be consistent
with the other clusters by taking its large errors into
account. This may suggest that the formation of the
gravitational potential well by the dark matter and the
metal injection process from galaxies to the ICM had been
already finished before $z\sim1.0$, but the gas accretion
process in which the primordial gas falls into the cluster
gravitational potential was going on at $z\sim1.0$. To
confirm it, we need the deep observations of X-ray clusters
at $z>1.0$, which will be obtained with forthcoming
observatories such as XMM, Chandra, ASTRO-E.  It may also be
possible that the ASCA results of AXJ2019 (Hattori et al.\
1997), particularly about its metallicity, may be in
error. For example, the detected iron line may come from a
foreground or background AGN. This will also be clarified by
future observations.

\section{Summary}

We analyzed the ASCA data for 29 nearby clusters ($z<0.1$)
and derived temperatures, iron abundances, and X-ray
luminosities. Furthermore, we fit the ASCA images, and then
determined the best-fit $\beta$ model parameters for the ICM
distribution. These results give the largest and most
homogeneous dataset about the ICM distribution obtained with
ASCA so far. We compare these results with distant clusters
whose temperatures, iron abundances, and luminosities were
also measured with ASCA. We found that there is no
significant difference between the clusters at $z<0.1$ and
$0.1<z<1.0$ in the $kT$ -- $L_{\rm X}$, $kT$ -- $M_{\rm
gas}$, $kT$ -- $A_{\rm Fe}$, $kT$ -- $M_{\rm Fe}$, and $kT$
-- $M_{\rm tot}$ relations. However, the most distant
cluster in our sample, AXJ2019 at $z=1.0$, may have
different characteristics; its ICM mass may be significantly
low, while its metallicity is quite large. They compensate
for each other and result in the iron mass which is similar
with the other clusters at $z<1.0$. This may suggest a hint
for the evolution of clusters of galaxies and that the
formation of the potential well and the metal injection
process of the ICM had finished before $z\sim1.0$, while the
accretion process of the primordial gas was going on at
$z\sim1.0$. However, it is also possible that AXJ2019 is
consistent with the other clusters taking its large errors
into account.

\par
\vspace{1pc}\par We would like to thank the ASCA team
members for their support. We are also grateful to K. Koyama
for helpful discussion and useful comments. HM is supported
by the JSPS Postdoctoral Fellowships for Research Abroad.

\clearpage
\section*{Reference}

\re
Anders E., Grevesse, N.\ 1989, Geochim. Cosmochim. Acta. 53, 197

\re
Donahue M., Vorr G. M., Gioia I., Luppino G., Hughes J. P.,
Stocke J. T.\ 1998, ApJ 502, 550

\re
Fukazawa Y.\ 1997, PhD Thesis, The University of Tokyo

\re
Fukazawa Y., Makishima K., Tamura T., Ezawa H., Xu H.,
Ikebe Y., Kikuchi K., Ohashi T.\ 1998, PASJ 50, 187

\re
Furuzawa A., Tawara Y., Kunieda H., Yamashita K., Sonobe T., 
Tanaka Y., Mushotzky R.\ 1998, ApJ 504, 35

\re
Hattori M., Ikebe Y., Asaoka I., Takeshima T., B\"ohringer H.,
Mihara T., Neumann D.\ M., Schindler S., Tsuru T., Tamura T.
1997, Nature 388, 146

\re
Hattori M., Matuzawa H., Morikawa K., Kneib J.-P., Yamashita K.,
Watanabe K., B\"ohringer H., Tsuru T.\ G.\ 1998, ApJ 503, 593

\re
Henry J.\ P., Henriksen M.\ J.\ 1986, ApJ 301, 689

\re
Hughes J.\ P., Birkinshaw M., Huchra J.\ P.\ 1995, ApJ 448, L93

\re
Markevitch M., Yamashita K., Furuzawa A., Tawara Y.\ 1994, ApJ 436, L71

\re
Markevitch M., Mushotzky R., Inoue H., Yamashita K., Furuzawa A.,
Tawara Y.\ 1996, ApJ 456, 437

\re
Masai K.\ 1984, Astrophys.\ Space Science 98, 367

\re
Matsuura M., Miyoshi S.\ J., Yamashita K., Tawara Y., Furuzawa A.,
Lasenby A.\ N., Saunders R., Jones M., Hatsukade I.\ 1996, ApJ 466, L75

\re
Mohr J.\ J., Mathiesen B., Evrard A.\ E.\ 1999, ApJ 517, 627

\re
Mushotzky R.\ F., Loewenstein M.\ 1997, ApJ 481, L63

\re
Mushotzky R.\ F., Scharf C.\ A.\ 1997, ApJ 482, L13

\re
Sarazin C.\ 1986, Rev. Mod. Phys. 58, 1

\re
Schindler S., Hattori M, Neumann D.\ M., B\"ohringer H.\ 1997, 
A\&A 317, 646

\re 
Takahashi T., Markevitch M., Fukazawa Y., Ikebe Y., Ishisaki
Y., Kikuchi K., Makishima K., Tawara Y., ASCA Image analysis
working group 1995, ASCA News No.3, p34

\re
Tanaka Y., Inoue H., Holt S.\ S.\ 1994, PASJ 46, L37

\re 
Tsuru T., Koyama K., Hughes J.\ P., Arimoto N., Kii T.,
Hattori M.\ 1996, in UV and X-ray Spectroscopy of
Astrophysical and Laboratory Plasmas, ed. K.\ Yamashita,
T.\ Watanabe (Universal Academy Press, Inc., Tokyo) p375

\re
White D.\ A., Jones C., Forman W. 1997, MNRAS 292, 419

\clearpage

\centerline{Figure Caption}
\bigskip

\begin{fv}{1}{}
{Temperature -- Luminosity relation. The luminosity is
measured in the 2 -- 10 keV band. Dots, and triangles denote
clusters at $z<0.1$, $0.1<z<1.0$, and AXJ2019 ($z=1.0$) is
denoted by a star.}
\end{fv}

\begin{fv}{2}{}
{Temperature -- ICM mass relation. The integration radius
for the ICM mass is 1 Mpc from the cluster center. The
symbols are the same as in figure 1.}
\end{fv}

\begin{fv}{3}{}
{Temperature -- iron abundance relation. The symbols are the
same as in figure 1.}
\end{fv}

\begin{fv}{4}{}
{Temperature -- iron mass relation. The integration radius
for the iron mass is 1.0 Mpc from the cluster center. The
symbols are the same as in figure 1.}
\end{fv}

\begin{fv}{5}{}
{Temperature -- total mass relation. The integration radius
for the total mass is 1.0 Mpc from the cluster center. The
symbols are the same as in figure 1.}
\end{fv}


\clearpage

\begin{table*}[t]
\begin{center}
Table~1.\hspace{4pt}Nearby cluster samples and results of
the spetral analysis.
\end{center}
\vspace{6pt}
\begin{tabular*}{\textwidth}{@{\hspace{\tabcolsep}
\extracolsep{\fill}}p{6pc}cccc}
\hline\hline\\[-6pt]
Name	&$L_{\rm X(2-10 keV)}^\dagger$	&$kT$	&$A_{\rm Fe}$	
&$N_{\rm H}$	\\
	&$10^{44}$erg/s	&keV	&solar	&$10^{20}$cm$^{-2}$	\\[4pt]\hline\\[-6pt]

Ophicuhus \dotfill	&20	&$10.0\pm1.5$	&$0.31\pm0.03$
&$28.0\pm1.1$	\\

A478 \dotfill	&16	&$6.40\pm0.25$	&$0.35\pm0.03$
&$20.9\pm3.7$	\\

A2319 \dotfill	&15	&$9.50\pm0.57$	&$0.23\pm0.05$
&$5.8\pm2.6$	\\

Tri Aust \dotfill	&12	&$9.86\pm0.57$	&$0.23\pm0.05$
&$14.6\pm2.5$	\\

Perseus \dotfill	&11	&$5.66\pm0.12$	& $0.43\pm0.02$
&$2.9\pm1.9$	\\

A1795 \dotfill		&8.2	&$5.68\pm0.11$	&$0.38\pm0.03$
&$0.0\pm3.0$	\\

Coma \dotfill		&7.6	&$8.95\pm0.25$	&$0.33\pm0.05$ 
&$0.0\pm0.30$	\\

A2256 \dotfill		&7.2	&$7.10\pm0.28$	&$0.28\pm0.04$
&$2.7\pm2.1$	\\

A85 \dotfill		&7.7	&$5.88\pm0.19$	&$0.43\pm0.04$ 
&$1.5\pm2.0$	\\

A3571 \dotfill		&6.6	&$7.24\pm0.24$	&$0.35\pm0.03$
&$3.0\pm1.7$	\\

A3558 \dotfill		&5.0	&$5.67\pm0.26$	&$0.29\pm0.05$
&$2.6\pm2.9$	\\

Hydra-A \dotfill	&3.3	&$3.71\pm0.14$	&$0.39\pm0.06$
&$0.6\pm3.0$	\\

A2199 \dotfill		&2.5	&$4.22\pm0.06$	&$0.38\pm0.04$
&$0.0\pm2.0$	\\

A496 \dotfill		&2.0	&$3.98\pm0.10$	&$0.45\pm0.04$
&$3.2\pm1.9$	\\

A119 \dotfill		&1.8	&$6.14\pm0.37$	&$0.35\pm0.07$
&$1.4\pm3.4$	\\

MKW3s \dotfill		&1.5	&$3.46\pm0.14$	&$0.38\pm0.07$
&$2.0\pm3.6$	\\

2A0335+096 \dotfill	&1.7	&$3.00\pm0.09$	&$0.48\pm0.06$
&$11.8\pm5.3$	\\

AWM7 \dotfill	&1.2	&$3.74\pm0.11$	&$0.55\pm0.06$ 
&$9.5\pm2.6$	\\
[4pt]\hline
\end{tabular*}
\end{table*}

\clearpage

\begin{table*}[t]
\begin{center}
Table~1.\hspace{4pt}Continued.
\end{center}
\vspace{6pt}
\begin{tabular*}{\textwidth}{@{\hspace{\tabcolsep}
\extracolsep{\fill}}p{6pc}cccc}
\hline\hline\\[-6pt]
Name	&$L_{\rm X(2-10 keV)}^\dagger$	&$kT$	&$A_{\rm Fe}$	
&$N_{\rm H}$	\\
	&$10^{44}$erg/s	&keV	&solar	&$10^{20}$cm$^{-2}$	\\[4pt]\hline\\[-6pt]

A2063 \dotfill		&1.0	&$3.72\pm0.11$	&$0.26\pm0.07$
&$0.0\pm3.5$	\\

A2147 \dotfill		&0.91	&$4.88\pm0.22$	&$0.36\pm0.06$
&$1.6\pm3.0$	\\

A2634 \dotfill		&0.52	&$3.58\pm0.19$	&$0.29\pm0.08$
&$1.0\pm4.2$	\\

Centaurus \dotfill	&0.53	&$3.52\pm0.09$	&$0.68\pm0.06$
&$3.5\pm2.3$	\\

A539 \dotfill		&0.42	&$3.27\pm0.16$	&$0.25\pm0.08$
&$6.7\pm4.3$	\\

A1060 \dotfill		&0.21	&$3.15\pm0.08$	&$0.43\pm0.05$
&$4.9\pm2.5$	\\

AWM4 \dotfill		&0.18	&$2.28\pm0.03$	&$0.33\pm0.13$
&$2.3\pm5.4$	\\

A400 \dotfill		&0.18	&$2.54\pm0.12$	&$0.33\pm0.10$ 
&$1.2\pm4.5$	\\

A262 \dotfill		&0.20	&$2.21\pm0.08$	&$0.33\pm0.10$
&$4.2\pm4.4$	\\

Virgo \dotfill		&0.16	&$2.28\pm0.03$	&$0.55\pm0.04$
&$7.2\pm2.8$	\\

MKW4s \dotfill		&0.065	&$2.19\pm0.21$	&$0.41\pm0.26$
&$2.1\pm9.7$	\\
[4pt]\hline
\end{tabular*}
\vspace{6pt}
\par\noindent
Errors are at 90\% confindence level.
\par\noindent
$\dagger$ Luminosity in the 2 -- 10 keV band.
\end{table*}

\clearpage

\begin{table*}[t]
\small
\begin{center}
Table~2.\hspace{4pt}Results of the imaging analysis.
\end{center}
\vspace{6pt}
\begin{tabular*}{\textwidth}{@{\hspace{\tabcolsep}
\extracolsep{\fill}}p{6pc}ccccccc}
\hline\hline\\[-6pt]
Name			&$\beta$	&\multicolumn{2}{c}{$r_{\rm core}$}	&$n(0)^{\dag}$	
&$M_{\rm gas}^\ddagger$	&$M_{\rm Fe}^\ddagger$	&$M_{\rm tot}^\ddagger$\\ \cline{3-4}
			&		&arcmin	&kpc	&$10^{-3}$cm$^{-3}$
&$10^{13}\MO$	&$10^{10}\MO$	&$10^{14}\MO$\\[4pt]\hline\\[-6pt]

Ophicuhus \dotfill	&$0.60\pm0.05$	&$3.50\pm0.50$	&227	&$8.0\pm1.6$
&14.6	&8.8	&6.4\\

A478 \dotfill		&$0.70\pm0.05$	&$1.00\pm0.25$	&154	&$17.0\pm3.0$
&11.6	&7.9	&4.9\\

A2319 \dotfill		&$0.55\pm0.05$	&$2.75\pm0.75$	&271	&$4.6\pm0.9$
&12.5	&5.6	&5.5\\

Tri Aust \dotfill	&$0.60\pm0.05$	&$3.00\pm0.50$	&252	&$5.1\pm1.0$
&10.9	&4.9	&6.3\\

Perseus \dotfill	&$0.45\pm0.05$	&$1.25\pm0.50$	&40	&$31.0\pm6.0$
&10.0	&8.4	&2.9\\

A1795 \dotfill		&$0.65\pm0.05$	&$1.25\pm0.25$	&135	&$13.0\pm3.0$
&8.5	&6.3	&4.1\\

Coma \dotfill		&$0.75\pm0.05$	&$10.25\pm0.50$	&416	&$2.8\pm0.6$ 
&8.9	&5.7	&6.5\\

A2256 \dotfill		&$0.75\pm0.05$	&$4.50\pm0.50$	&457	&$2.6\pm0.5$
&9.4	&5.2	&5.0\\

A85 \dotfill		&$0.60\pm0.05$	&$1.75\pm0.50$	&173	&$7.4\pm1.5$ 
&8.9	&7.5	&3.9\\

A3571 \dotfill		&$0.60\pm0.05$	&$2.50\pm0.50$	&171	&$6.7\pm1.3$
&7.9	&5.4	&4.8\\

A3558 \dotfill		&$0.50\pm0.05$	&$1.75\pm0.50$	&150	&$5.2\pm1.0$
&7.5	&4.3	&3.1\\

Hydra-A \dotfill	&$0.60\pm0.05$	&$0.75\pm0.50$	&71	&$20.0\pm8.0$
&5.5	&4.2	&2.5\\

A2199 \dotfill		&$0.60\pm0.05$	&$1.50\pm0.50$	&80	&$14.0\pm3.0$
&4.8	&3.5	&2.8\\

A496 \dotfill		&$0.55\pm0.05$	&$1.25\pm0.50$	&70	&$13.0\pm3.0$
&4.7	&4.2	&2.5\\

A119 \dotfill		&$0.60\pm0.05$	&$6.00\pm0.50$	&462	&$0.98\pm0.20$
&4.6	&3.2	&3.4\\

MKW3s \dotfill		&$0.65\pm0.05$	&$1.25\pm0.30$	&95	&$11.0\pm2.0$
&3.9	&2.9	&2.5\\

2A0335+096 \dotfill	&$0.60\pm0.05$	&$0.75\pm0.25$	&46	&$30.0\pm6.0$
&3.9	&3.7	&2.0\\

AWM7$^*$ \dotfill	&$0.55\pm0.05$	&$4.00\pm1.00$	&125	&$4.6\pm0.9$
&$\cdots$	&$\cdots$	&$\cdots$\\
[4pt]\hline
\end{tabular*}
\end{table*}

\clearpage

\begin{table*}[t]
\small
\begin{center}
Table~2.\hspace{4pt}Continued.
\end{center}
\vspace{6pt}
\begin{tabular*}{\textwidth}{@{\hspace{\tabcolsep}
\extracolsep{\fill}}p{6pc}ccccccc}
\hline\hline\\[-6pt]
Name			&$\beta$	&\multicolumn{2}{c}{$r_{\rm core}$}	&$n(0)^{\dag}$	
&$M_{\rm gas}^\ddagger$	&$M_{\rm Fe}^\ddagger$	&$M_{\rm tot}^\ddagger$\\ \cline{3-4}
			&		&arcmin	&kpc	&$10^{-3}$cm$^{-3}$
&$10^{13}\MO$	&$10^{10}\MO$	&$10^{14}\MO$\\[4pt]\hline\\[-6pt]

A2063 \dotfill		&$0.60\pm0.05$	&$2.25\pm0.50$	&133	&$4.8\pm1.0$
&3.8	&1.9	&2.5	\\

A2147 \dotfill		&$0.50\pm0.05$	&$3.50\pm0.50$	&218	&$1.5\pm0.3$
&3.6	&2.5	&2.6\\

A2634 \dotfill		&$0.50\pm0.05$	&$4.50\pm1.20$	&246	&$1.1\pm0.2$
&3.0	&1.7	&1.9\\

Centaurus$^*$ \dotfill	&$0.50\pm0.05$	&$4.00\pm0.50$	&75	&$5.5\pm1.1$
&$\cdots$	&$\cdots$	&$\cdots$\\

A539 \dotfill		&$0.60\pm0.05$	&$3.50\pm0.50$	&164	&$2.5\pm0.5$
&2.8	&1.4	&2.2\\

A1060$^*$ \dotfill	&$0.55\pm0.05$	&$4.00\pm1.00$	&80	&$4.4\pm0.9$
&$\cdots$	&$\cdots$	&$\cdots$\\

AWM4 \dotfill		&$0.55\pm0.05$	&$1.25\pm1.20$	&68	&$5.1\pm2.1$
&1.8	&1.1	&1.4\\

A400$^*$ \dotfill	&$0.45\pm0.05$	&$2.50\pm0.80$	&102	&$2.1\pm0.4$
&$\cdots$	&$\cdots$	&$\cdots$\\

A262$^*$ \dotfill	&$0.50\pm0.05$	&$2.25\pm0.75$	&63	&$5.5\pm1.1$
&$\cdots$	&$\cdots$	&$\cdots$\\

Virgo$^*$ \dotfill	&$0.40\pm0.05$	&$2.00\pm1.00$	&14	&$19.0\pm8.0$
&$\cdots$	&$\cdots$	&$\cdots$\\

MKW4s$^*$ \dotfill	&$0.40\pm0.05$	&$1.00\pm0.50$	&47	&$2.7\pm1.1$
&$\cdots$	&$\cdots$	&$\cdots$\\
[4pt]\hline
\end{tabular*}
\vspace{6pt}
\par\noindent
Errors are at 90\% confindence level.
\par\noindent
$\dag$ The central proton density.
\par\noindent
$\ddagger$ Mass within th radius of 1 Mpc from cluster center.
\par\noindent
$*$ Gas and iron masses are not listed hear, because the FOV
of GIS can covers only small region with radii smaller than
1 Mpc from cluster center.
\end{table*}

\clearpage

\begin{table*}[t]
\scriptsize
\begin{center}
Table~3.\hspace{4pt}Distant cluster samples
\end{center}
\vspace{6pt}
\begin{tabular*}{\textwidth}{@{\hspace{\tabcolsep}
\extracolsep{\fill}}p{5pc}cccccccc}
\hline\hline\\[-6pt]
Name		&$z$	&$L_{\rm X(2-10 keV)}^\dagger$	&$kT$	&$A_{\rm Fe}$
&$M_{\rm gas}^{\ddag}$	&$M_{\rm Fe}^{\ddag}$	&$M_{\rm tot}^{\ddag}$	&Reference$^*$\\
&	&$10^{44}$erg/s	&keV	&solar	&$10^{13}\MO$	&$10^{10}\MO$
&$10^{14}\MO$\\[4pt]\hline\\[-6pt]

A1413 \dotfill	&0.1430	&13	&$6.72\pm0.26$	&$0.29\pm0.05$	
&7.4		&4.2	&8.9	&j, k, m\\

A2204 \dotfill	&0.1530	&33	&$8.47\pm0.42$	&$\cdots$	
&$\cdots$	&$\cdots$	&$\cdots$	&j\\

A1204 \dotfill	&0.1700	&6.3	&$3.83\pm0.19$	&$0.35\pm0.07$	
&7.0		&4.8	&3.0	&j, k, i\\

A2218 \dotfill	&0.1710	&9.4	&$7.04\pm0.97$	&$0.18\pm0.07$	
&$\cdots$	&$\cdots$	&$\cdots$	&j, k\\

A586 \dotfill	&0.1710	&8.3	&$6.61\pm1.15$	&$\cdots$	
&$\cdots$	&$\cdots$	&$\cdots$	&j\\

A1689 \dotfill	&0.1800	&30	&$9.02\pm0.40$	&$0.26\pm0.06$	
&12.7		&6.5	&9.6	&j, k, m\\

A1246 \dotfill	&0.1870	&8.3	&$6.28\pm0.54$	&$0.22\pm0.08$	
&7.8		&3.4	&5.1	&j, k, m\\

A1763 \dotfill	&0.1870	&15	&$8.98\pm1.02$	&$0.26\pm0.09$
&10.7		&5.4	&5.6	&j, k, m\\

MS0440 \dotfill	&0.1900	&3.9	&$5.3\pm1.3$	&$\cdots$
&$\cdots$	&$\cdots$	&$\cdots$	&j\\

MS0839 \dotfill	&0.1940	&4.0	&$4.19\pm0.36$	&$\cdots$
&$\cdots$	&$\cdots$	&$\cdots$	&j\\

A773 \dotfill	&0.1970	&13	&$9.66\pm1.03$	&$0.24\pm0.08$
&$\cdots$	&$\cdots$	&$\cdots$	&j, k\\

A520 \dotfill	&0.2010	&16	&$8.59\pm0.93$	&$0.25\pm0.20$
&$\cdots$	&$\cdots$	&$\cdots$	&j, k\\

A2163 \dotfill	&0.2010	&16	&$12.7\pm2.0$	&$0.38\pm0.13$	
&14.9		&11.6	&12.6	&j, h, g, m\\

A963 \dotfill	&0.2060	&8.9	&$6.76\pm0.44$	&$0.29\pm0.08$
&$\cdots$	&$\cdots$	&$\cdots$	&j, k\\

A1704 \dotfill	&0.2190	&6.3	&$4.51\pm0.56$	&$\cdots$
&$\cdots$	&$\cdots$	&$\cdots$	&j\\

A2219 \dotfill	&0.2280	&33	&$11.77\pm1.26$	&$0.25\pm0.07$
&$\cdots$	&$\cdots$	&$\cdots$	&j, k\\

A2390 \dotfill	&0.2300	&22	&$8.90\pm0.97$	&$0.22\pm0.06$
&11.2		&4.8	&8.3	&j, k, m\\

MS1305+29 \dotfill	&0.2410	&0.89	&$2.98\pm0.52$	&$\cdots$
&$\cdots$	&$\cdots$	&$\cdots$	&j\\

A1835 \dotfill	&0.2520	&45	&$8.15\pm0.46$	&$0.32\pm0.05$
&$\cdots$	&$\cdots$	&$\cdots$	&j, k\\

MS1455 \dotfill	&0.2580	&12	&$5.45\pm0.29$	&$0.33\pm0.08$
&$\cdots$	&$\cdots$	&$\cdots$	&j, k\\

A1758N \dotfill	&0.2800	&17	&$10.19\pm2.29$	&$\cdots$
&$\cdots$	&$\cdots$	&$\cdots$	&j\\

A483 \dotfill	&0.2830	&3.3	&$6.87\pm1.59$	&$\cdots$
&$\cdots$	&$\cdots$	&$\cdots$	&j\\

ZW3146 \dotfill	&0.2900	&27	&$6.35\pm0.37$	&$0.24\pm0.05$
&$\cdots$	&$\cdots$	&$\cdots$	&j, k\\

MS1008-12 \dotfill	&0.3010	&6.9	&$7.29\pm2.45$	&$\cdots$
&$\cdots$	&$\cdots$	&$\cdots$	&j\\

AC118 \dotfill	&0.3080	&25	&$12.08\pm1.42$	&$0.23\pm0.09$
&$\cdots$	&$\cdots$	&$\cdots$	&j, k\\

MS2137 \dotfill	&0.3130	&10	&$4.37\pm0.38$	&$0.41\pm0.12$
&$\cdots$	&$\cdots$	&$\cdots$	&j, k\\

A1995 \dotfill	&0.3180	&13	&$10.70\pm2.50$	&$\cdots$
&$\cdots$	&$\cdots$	&$\cdots$	&j\\

MS0353-36 \dotfill	&0.3200	&7.5	&$8.13\pm2.57$	&$\cdots$
&$\cdots$	&$\cdots$	&$\cdots$	&j\\
[4pt]\hline
\end{tabular*}
\end{table*}

\clearpage

\begin{table*}[t]
\scriptsize
\begin{center}
Table~3.\hspace{4pt} Continued.
\end{center}
\vspace{6pt}
\begin{tabular*}{\textwidth}{@{\hspace{\tabcolsep}
\extracolsep{\fill}}p{5pc}cccccccc}\hline\hline
Name		&$z$	&$L_{\rm X(2-10 keV)}^\dagger$	&$kT$	&$A_{\rm Fe}$
&$M_{\rm gas}^{\ddag}$	&$M_{\rm Fe}^{\ddag}$	&$M_{\rm tot}^{\ddag}$	&Reference$^*$\\
&	&$10^{44}$erg/s	&keV	&solar	&$10^{13}\MO$	&$10^{10}\MO$
&$10^{14}\MO$\\[4pt]\hline\\[-6pt]

A1722 \dotfill	&0.3270	&8.3	&$5.87\pm0.51$	&$0.25\pm0.11$
&$\cdots$	&$\cdots$	&$\cdots$	&j, k\\

MS1358 \dotfill	&0.3270	&7.6	&$6.50\pm0.68$	&$0.27\pm0.10$
&$\cdots$	&$\cdots$	&$\cdots$	&j, k\\

A959 \dotfill	&0.3530	&11	&$6.95\pm1.85$	&$\cdots$
&$\cdots$	&$\cdots$	&$\cdots$	&j\\

MS1512+36 \dotfill	&0.3720	&3.7	&$3.57\pm1.33$	&$\cdots$
&$\cdots$	&$\cdots$	&$\cdots$	&j\\

A370 \dotfill	&0.3730	&14	&$7.13\pm1.05$	&$\cdots$
&$\cdots$	&$\cdots$	&$\cdots$	&j\\

A851 \dotfill	&0.4100	&5.8	&$6.7\pm2.7$	&$\cdots$
&$\cdots$	&$\cdots$	&$\cdots$	&j\\

RXJ1347-114 \dotfill	&0.4510	&88	&$11.37\pm1.10$	&$0.33\pm0.10$
&20.0		&12.9	&5.8	&j, k, l\\

3C295 \dotfill	&0.4600	&9.1	&$7.13\pm2.06$	&$\cdots$
&5.3		&$\cdots$	&5.7	&j, e\\

MS0451-03 \dotfill	&0.5390	&19	&$10.17\pm1.55$	&$0.16\pm0.12$
&$\cdots$	&$\cdots$	&$\cdots$	&j, k\\

CL0016 \dotfill	&0.5410	&14	&$8.0\pm1.0$	&$0.11\pm0.12$
&$\cdots$	&$\cdots$	&6.1	&j, b, f\\

CL2236-94 \dotfill	&0.552	&4.9	&$6.1\pm2.6$	&$0.00(<0.38)$
&5.8		&$0.0(<4.3)$	&4.4	&d\\

MS1054 \dotfill	&0.829	&23	&$12.3\pm3.1$	&$0.00(<0.25)$
&$\cdots$	&$\cdots$	&$\cdots$	&a\\

AXJ2019+1127 \dotfill	&1.00	&8.4	&$8.6\pm{+4.2}$	&$1.7^{+1.3}_{-0.7}$
&2.4		&8.0	&3.6	&c\\
			&	&	&		&
&(4.5)$^{\S}$	&$(15.1)^{\S}$	&(8.5)$^{\S}$\\
[4pt]\hline
\end{tabular*}
\vspace{6pt}
\par\noindent
Errors are at 90\% confindence level except for AXJ2019.
The errors for AXJ2019 are at 1 $\sigma$ level.
\par\noindent
$\dagger$ Luminosity in the 2 -- 10 keV band.
\par\noindent
$\ddag$ Mass within the radius of 1 Mpc from the cluster center.
\par\noindent
$\S$ The ICM in AXJ2019 was detected only within 0.5 Mpc
from the center.  The values in parenthesis are masses
within 1.0 Mpc estimated by extrapolating the best-fit
$\beta$ model.
\par\noindent
$*$References; 
a. Donahue et al.\ (1998),
b. Furuzawa et al.\ (1998), 
c. Hattori et al.\ (1997), 
d. Hattori et al.\ (1998), 
e. Henry, Henriksen (1986),
f. Hughes, Birkinshaw, Huchra (1995)
g. Markevitch et al.\ (1994),
h. Markevich et al.\ (1996), 
i. Matsuura et al.\ (1996),
j. Mushotzy, Scharf (1997), 
k. Mushotzy, Loewenstein (1997), 
l. Schindler et al.\ (1997), 
m. White et al.\ (1997), 
\end{table*}

\clearpage

\input epsf
fig1\\
\epsfsize = 0.9\textwidth
\epsfbox[25 18 587 774]{figure1.eps}

\clearpage
fig2\\
\epsfsize = 0.9\textwidth
\epsfbox[25 18 587 774]{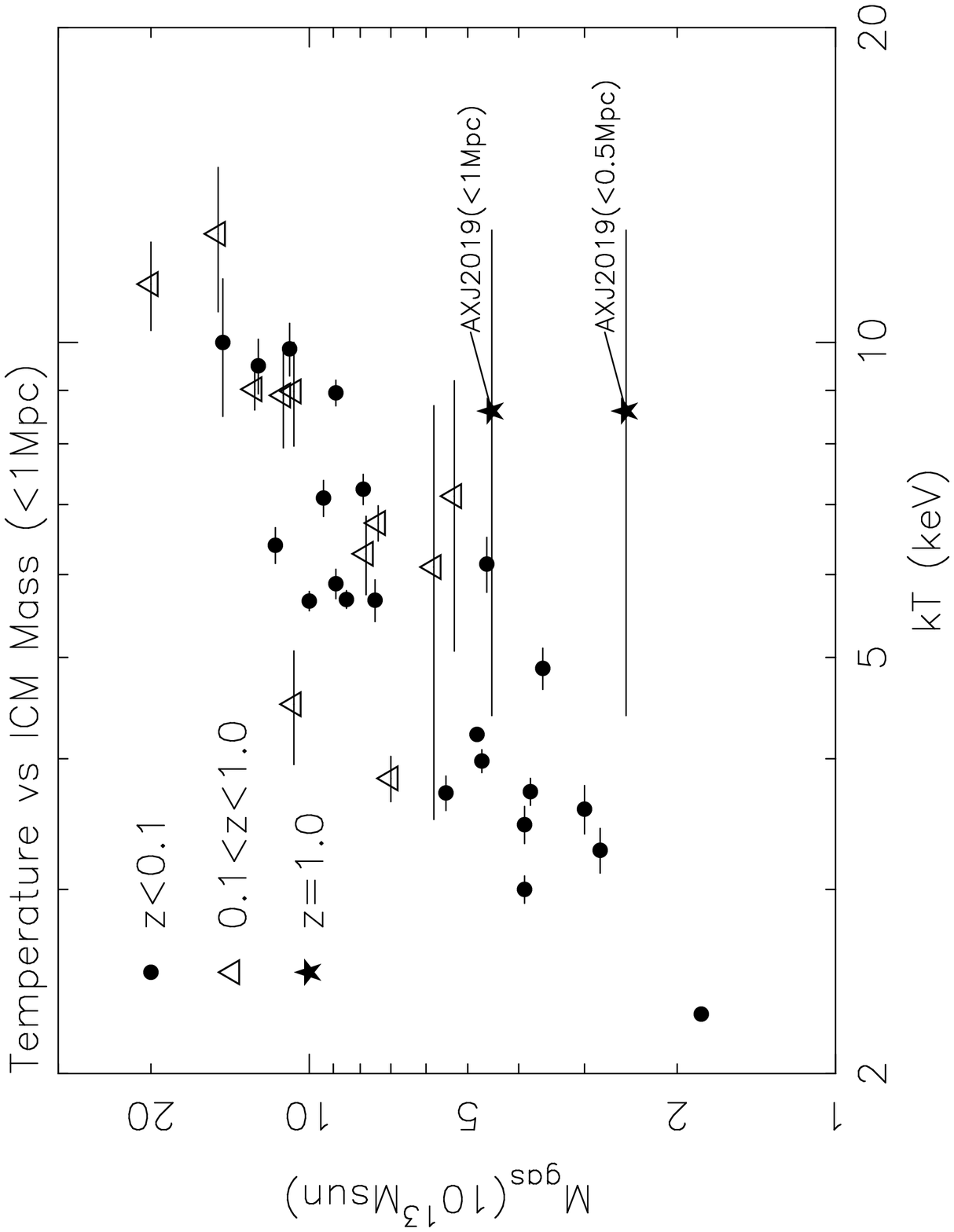}

\clearpage
fig3\\
\epsfsize = 0.9\textwidth
\epsfbox[25 18 587 774]{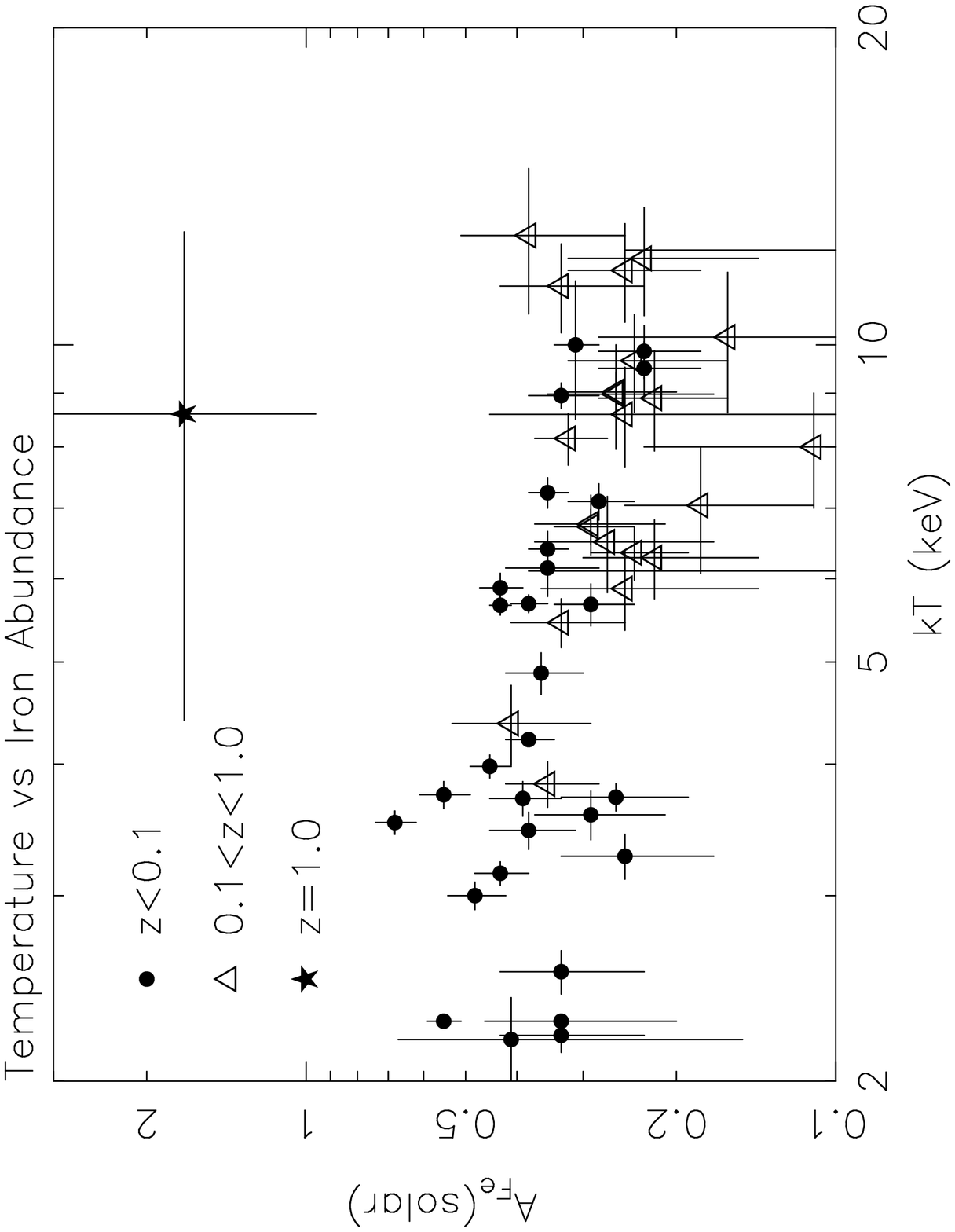}

\clearpage
fig4\\
\epsfsize = 0.9\textwidth
\epsfbox[25 18 587 774]{figure4.eps}

\clearpage
fig5\\
\epsfsize = 0.9\textwidth
\epsfbox[25 18 587 774]{figure5.eps}

\end{document}